\documentclass[twocolumn,showpacs,preprintnumbers,amsmath,amssymb]{revtex4}

\usepackage{graphicx}
\usepackage{dcolumn}
\usepackage{bm}

\begin{document}

\preprint{APS/   }

\title{Green function Retrieval and Time-reversal in a Disordered World}

\author{B.A. van Tiggelen}
\affiliation{%
Laboratoire de Physique et Mod\'elisation des Milieux
Condens\'es,CNRS/Maison de Magist\`eres\\ Universit\'e Joseph
Fourier
B.P. 166, 38042 Grenoble Cedex 9, France\\
}%

\date{\today}

\begin{abstract}
We apply the theory of multiple wave scattering to two
contemporary, related topics: imaging with diffuse correlations
and stability of time-reversal of diffuse waves, using
equipartition, coherent backscattering and frequency speckles as
fundamental concepts.
\end{abstract}

\maketitle

In its early days, multiple scattering of waves was considered to
be an unavoidable nuisance. It randomizes phase, polarization and
wave vector of waves, and thus complicates important applications
in imaging, telecommunication, laser action, and remote sensing.
Recent developments  showed that multiple scattering can actually
\emph{enhance} the performance of several applications. We mention
in this context the low threshold of random lasers \cite{rl}, the
robust time-reversal of multiply scattered waves \cite{derode95},
with potential applications in communication \cite{tourin}, the
sensitivity of diffuse waves to particle motions \cite{dws}, and
the retrievals of the Green function from thermal phonons
\cite{weaver} and diffuse seismic waves \cite{anne}.

Chaos theory successfully describes time-reversal
\cite{julienthese} and correlations \cite{weaver} of elastic waves
in chaotic media. Most applications above concern diffuse waves,
with their specific questions about statistics, leakage and
dynamics. It is the intention of this Letter to apply multiple
scattering theory to these new exciting topics. To our knowledge
the first attempts to cross-correlate "noisy" signals to retrieve
ballistic wave motion come from seismology ("acoustic daylight
imaging") \cite{ricket} and helio-seismology \cite{helio}. As for
time-reversal, after many pioneering experiments in Paris
\cite{finkall}, Papanicolaou \emph{etal}. \cite{george2} explained
the self-averaging property of time-reversal for broadband signals
for (locally) layered random media in a scaling limit of small
fluctuations and propagation distances long compared to the
wavelength. Our approach makes no assumptions on the scatterers,
but applies only in the diffuse regime, i.e. when the medium size
$L$ is much bigger than the mean free path $\ell^*$. This excludes
1D and localized media. This criterion is largely fulfilled in 2D
time-reversal experiments with high order multiple scattering
\cite{derode95}.

We first consider the fluctuations of a scalar wave field
$\Psi(\textbf{r},t)$ propagating in an infinite random medium
after being released by some source $S$ far away from the place of
measurement. It is customary to describe field correlations by the
"Wigner" function \cite{papa},
$\Psi\left(\textbf{r}-\frac{1}{2}\textbf{x},
t-\frac{1}{2}\tau\right)\,
\Psi\left(\textbf{r}+\frac{1}{2}\textbf{x},t+\frac{1}{2}\tau\right)$,
whose Fourier transform with respect to space $\textbf{x}$ and
time $\tau$ is defined as the specific intensity
$I_{\omega\textbf{k}}\left( \textbf{r},t\right)$ of waves with
wave vector $\textbf{k}$ and frequency $\omega$. A \emph{rigorous}
statement in multiple scattering theory is that
\begin{eqnarray}\label{ward}
    \lim_{t\rightarrow \infty}\, \int \textrm{d}\textbf{r}
    \langle I_{\omega\textbf{k}}\left(\textbf{r},t\right) \rangle = -\frac{1}{\omega}
    \textrm{Im}\,
    \langle G(\omega,\textbf{k})\rangle  \nonumber \\ \times \ \frac{\sum_{\textbf{k'}}-\frac{1}{\omega}\textrm{Im}\,
    \langle G(\omega,\textbf{k'})\rangle S(\omega, \textbf{k'})}{\sum_{\textbf{k'}}-\frac{1}{\omega}\textrm{Im}\,
    \langle G(\omega,\textbf{k'}) \rangle}\  .
\end{eqnarray}
The structure  function of the source  $S(\omega, \textbf{k})$ is
 the FT of $S(\tau,\textbf{x}) \equiv \int \textrm{d}t\int
\textrm{d}\textbf{r} \, s\left(\textbf{r}-\frac{1}{2}\textbf{x},
t-\frac{1}{2}\tau\right)\,
s\left(\textbf{r}+\frac{1}{2}\textbf{x},t+\frac{1}{2}\tau\right)$,
and $\langle G(\omega,\textbf{k})\rangle$ is the  \emph{average}
Green's function of the random medium. Equation~(\ref{ward}) is a
known field-theoretical consequence of flux conservation
\cite{pr,barbie}. It relates the specific intensity \emph{at large
lapse times} to the spectral function $-\frac{1}{\omega}
\textrm{Im}\, \langle G(\omega,\textbf{k}) \rangle $ of the
effective medium \cite{ping}. This implies global equipartition of
average energy in phase space.

Equation~(\ref{ward}) looks like a manifestation of the
fluctuation-dissipation (FD) theorem, applied to $\Psi = G \otimes
s$ for which thermal equilibrium (TE) implies that
$\left<\Psi(\omega)\Psi^*(\omega)\right> \propto \textrm{Im}\,
\langle G (\omega) \rangle B(\omega)$, with $B$ the Planck
function.  In TE, formula~(\ref{ward}) would even hold without
multiple scattering, and at any time. This situation applies to
the experiment by Weaver and Lobkis \cite{weaver}, who measured
the elastic Green's function of an aluminium block by
cross-correlating thermal phonons. When translated to space-time,
the FD theorem shows that coherent wave paths should in principle
be observable with travel times up to the inelastic phonon mean
free time.

The diffuse field is not in TE, not even in equilibrium. The
physics that multiple scattering and TE share is the equipartition
principle \cite{richard,papa}. Unfortunately, two reasons exist
why Eq.~(\ref{ward}) is not very useful to retrieve the Green
function. First, it applies to the ensemble-average of the
specific intensity only, which is subject to large statistical
fluctuations. This would clearly restrict  its  use in several
imaging problems.
    Secondly, Eq.~(\ref{ward}) assumes  many detectors at different
    positions. As
shown in Ref. \cite{weaver}, the correlation method can  be
adapted for infinitely many equal sources, which also eliminates
the problem of ensemble-averaging. In this work we will consider
the random field generated by one distant source and investigate
the applicability to retrieve the Green's function using the
\emph{local} time correlation function
$\Phi_{\textbf{r}}(\textbf{x},\tau) \equiv \int \textrm{d}t \,
\Psi\left(\textbf{r}-\frac{1}{2}\textbf{x},
t-\frac{1}{2}\tau\right)\,
\Psi\left(\textbf{r}+\frac{1}{2}\textbf{x},t+\frac{1}{2}\tau\right)$.
We shall
    develop a theory for its  \emph{ statistical average}, and show
    that $\Phi_{\textbf{r}}(\textbf{x},\tau)$ tends to be self-averaging.

Equation~(\ref{ward}) suggests the following local expansion of
the ensemble-averaged  specific intensity,
\begin{eqnarray}\label{da}
     \langle I_{\omega\textbf{k}}(\textbf{r},t) \rangle= -\frac{1}{\omega}
    \textrm{Im}\,
    \langle G(\omega,\textbf{k}) \rangle \left[ 1 - v_{p} \ell^*
    \frac{1}{\omega}\textbf{k}\cdot \partial_{\textbf{r}} + \cdots \right]
    \nonumber \\
    \times S(\omega)\rho_{\omega}(\textbf{r},t) \, .
\end{eqnarray}
where the energy density  $\rho_{\omega}(\textbf{r},t)$ satisfies
a
 diffusion equation with source
$\delta(\textbf{r})\delta(t)$, and the power spectrum $S(\omega)$
is defined as the source factor on the righthand side of
Eq.~(\ref{ward}). The second term in Eq.~(\ref{da}) gives a
diffuse flow of energy and is known to be valid in dimensions
$d>1$.
 Let us
consider an ideal explosive source
$s(\textbf{r},\tau)=S\delta(t)\delta'(\textbf{r})$, with power
spectrum  $S(\omega) = S \omega^{2}$ that we band-filter in a
frequency band $B$, inside which transport quantities like the
diffusion constant $D$ are assumed not to vary too much. We get
the simple result,
\begin{eqnarray}\label{corr2}
\left< \Phi_{\textbf{r}}(\textbf{x},\tau)\right> = S
\left[\rho(\textbf{r})\partial_{\tau} - v_{p}\ell^*
\partial_{\textbf{r}}\rho(\textbf{r})\cdot
\partial_{\textbf{x}} + \cdots \right] \, \times \nonumber \\
 \left[ \langle G_{B}(\textbf{x},\tau) \rangle - \langle G_{B}(\textbf{x},-\tau)
 \rangle \right].
\end{eqnarray}
The ensemble-averaged field correlation function is proportional
to the time-derivative of the band-filtered, ensemble-averaged
Green's function $\langle G_{B}\rangle $ of the random medium,
\emph{symmetrized} in the time $\tau$ \cite{weaver}. An
\emph{antisymmetric} part is allowed if a diffuse flux is present.
There is a strong belief that this term is the origin of the
asymmetry of the correlation function recently observed with
seismic Rayleigh waves \cite{anne}. The verification of this
hypothesis requires a more realistic diffusion model for the Earth
crust, including (free) surface detection, surface waves and mode
conversions \cite{nicolaspre}, that is beyond the scope of this
Letter, and for which laboratory experiments will be indispensable
\cite{alison}. The diffuse character of these seismic waves had
been deduced earlier from their  equipartition \cite{prlequi} and
their seismic coda  \cite{shapiro}. In an open medium of size $L$
the relative importance of the asymmetric term is of order
$\ell^*/L$
 far away from its  boundaries, and thus small if $L\gg \ell^*$.

If the waves propagate in 3D without much dispersion, $\langle
G(\textbf{x},\tau)\rangle   \approx -\delta(x-c\tau)/4\pi x$ for
$\tau$ small compared to the mean free time, and Eq.~(\ref{corr2})
can easily be generalized for an \emph{arbitrary} source spectrum,
\begin{equation}\label{nodisp}
\partial_{\tau}\left<\Phi_{\textbf{r}}(\textbf{x},\tau)\right>
\rightarrow   \frac{\rho(\textbf{r})}{4\pi x}
\left[S\left(\frac{x}{c} + \tau\right) -
S\left(\frac{x}{c}-\tau\right) \right] \, .
\end{equation}
$S(\tau)$ is the time correlation function of the source. In
particular, the result $\left<\Phi_{\textbf{r}}(0,\tau)\right>
\sim \rho(\textbf{r}) S(\tau)$ will be needed later to describe
time-reversal. Relation~(\ref{nodisp}) facilitates the monitoring
of source dynamics with distant, diffuse correlations.

Equations~(\ref{corr2},\ref{nodisp}) constitute the central result
of this paper, despite their simplicity. Their importance follows
from the rest of this Letter. First we establish that fluctuations
around the ensemble-average are small if the bandwidth $B$ is
large enough, as also found for layered random media
\cite{george2, george1}, and first suggested by time-reversal
experiments \cite{selfav}. We will generalize the central
result~(\ref{corr2}) to a random medium containing an
\emph{arbitrary} close object. This shows the  usefulness of
diffuse waves to the inverse problem in disordered media, in a way
as first put forward by Claerbout \emph{etal}. \cite{ricket} for
seismic noise. At the end of this Letter we make the link with
time-reversal and coherent backscattering in disordered media
\cite{finkall}.

We start out by calculating the statistical fluctuations of
$\Phi_{\textbf{r}}(\textbf{x},\tau)$ around its average. To this
end let us adopt a source at a large distance $\textbf{r}$ with
power spectrum $S(\omega)= S \omega^{2}$ that we filter over a
bandwidth $B$. Our most important assumption will be that the
bandwidth $B$ is much larger than the Thouless frequency
$\Omega_{\textrm{th}}= D/2d|\textbf{r}|^2$ of the random medium
(with $D$ the diffusion constant, and $d$ the dimension). It can
readily be seen that ($\textbf{r}^{\pm} = \textbf{r}
\pm\frac{1}{2}\textbf{x}$),
\begin{eqnarray}\label{inter1}
\Phi_{\textbf{r}}(\textbf{x},\tau) =  \int_{-\infty}^{\infty}
\frac{\textrm{d} \omega}{2\pi}  \textrm{e}^{-i\omega\tau}\,
G(\textbf{r}^{-} , 0,\omega)G^*(\textbf{r}^{+}, 0, \omega )
  S(\omega)\, .
\end{eqnarray}
in terms of the monochromatic, retarded Green's function
$G(\textbf{r},\textbf{r}',\omega)$ of the wave equation. From
Eq.~(\ref{da})  we find for the average correlation,
\begin{eqnarray}\label{inter2}
\langle\Phi_{\textbf{r}}(0,0)\rangle = \int_{-\infty}^{\infty}
\frac{\textrm{d} \omega}{2\pi}\, \frac{-1}{\omega}\, \textrm{Im}\,
\langle G(\omega, 0) \rangle S \rho_{\omega}(\textbf{r})  \simeq B
N S \rho(\textbf{r}) \nonumber
\end{eqnarray}
with $N$ the density of states in the bandwidth \cite{pr}. In the
diffuse regime spatial/frequency fluctuations obey Gaussian
($C_{1}$) statistics \cite{azi}, so that the variance becomes,
\begin{widetext}
\begin{eqnarray}\label{inter3}
\left(\Delta \Phi_{\textbf{r}}(\textbf{x},\tau) \right)^2 &=&
\int_{-\infty}^{\infty} \frac{\textrm{d} \omega_{1}}{2\pi}
\int_{-\infty}^{\infty} \frac{\textrm{d} \omega_{2}}{2\pi}
\,\textrm{e}^{-i(\omega_{1}-\omega_{2})\tau }
S(\omega_{1})S(\omega_{2}) \left< G(\textbf{r}^{-} , 0,\omega
_{1})  G^*(\textbf{r}^{-} , 0,\omega_{2}) \right>
\left<G(\textbf{r}^{+} , 0,\omega _{2})\, G^*(\textbf{r}^{+},
0,\omega_{1}) \right> \nonumber
\\ &\approx& \int_{B} \frac{\textrm{d} \omega_{1}}{2\pi}\, S^{2}
N(\omega_{1})^2\int_{-\infty}^{\infty} \textrm{d} \Omega \,
\textrm{e}^{-i\Omega\tau} \rho_{\omega}(\textbf{r})^2
\left|C^{1}(\Omega)\right|^2 \nonumber \\
 &\simeq&   B  |S|^2N^{2}
 \rho(\textbf{r})^2 \Omega_{\textrm{th}}\, .\nonumber
\end{eqnarray}
\end{widetext}
We use  that the normalized intensity correlation function
$C^{1}(\Omega) \thicksim
\exp(-\sqrt{\Omega/\Omega_{\textrm{th}}})$ decays rapidly with
$\Omega$ and that $\Omega_{\textrm{th}} \ll B, 1/\tau$. In that
case is $ \Delta \Phi_{\textbf{r}}/\langle\Phi_{\textbf{r}}\rangle
    \simeq \sqrt{\Omega_{\textrm{th}}/{B}} \ll 1
$, so that the measurement of $\Phi_{\textbf{r}}(\textbf{x},\tau)$
is close to its ensemble-average with high probability. The ratio
$B/\Omega_{\textrm{th}}$ is interpreted as the number of
independent "frequency bits" available in the bandwidth of the
waves arriving at the receiver \cite{selfav}.

In the following we will show the possibility to \emph{image} a
fixed, close object by means of diffuse correlations of a scalar
field $\Psi$. The formal relation between the Green function
$G^{T}$ of the effective medium \emph{including} the object, the
Green function $\langle G\rangle$ of the effective medium alone,
and the T-matrix $T$ of the object is $G^{T} = G + GTG$ (we drop
the averaging brackets if no confusion can arise). We assert that
Eq.~(\ref{corr2}) is universal so that at a large distance
$\textbf{r}$ from an explosive source,
\begin{eqnarray}\label{object}
\Phi_{\textbf{r}}(\textbf{x},\tau)
  &&\propto
\rho(\textbf{r})\nonumber \\
&&\partial_{\tau} \left[ \langle
G^{T}_{B}(\textbf{r}^{-},\textbf{r}^{+},\tau) \rangle - \langle
G^{T}_{B}(\textbf{r}^{-},\textbf{r}^{+},-\tau) \rangle \right].
\end{eqnarray}
where $B$ refers again to a finite bandwidth. The
identity~(\ref{object}) implies that the field correlation of the
diffuse waves  between two points  near the object is equivalent
to a time-resolved scattering experiment on the object.

This assertion can be established from the following assumptions:
1. We are in the diffuse field of a distant (explosive) source
$S$; 2. the object scatters the waves elastically. Our starting
point is Eq.~(\ref{da}) valid without the object. The solution
\emph{with} object can formally be found by "gluing" a single
scattering vertex $K$ into the transport vertex $\rho$ giving the
terms $(\rho + K\rho + \rho K + \rho K\rho)S$, with $S$ the
source. In momentum space the vertex $K$ at frequency $\omega$ is
given by \cite{NvR},
\begin{eqnarray}\label{K}
     K_{\textbf{kk'}}(\textbf{q}_{1}, \textbf{q}_{2}) &=&
    T_{\textbf{k}+\frac{1}{2}\textbf{q}_{1}, \textbf{k'}+\frac{1}{2}\textbf{q}_{2}}
     T^{*}_{\textbf{k}-\frac{1}{2}\textbf{q}_{1}, \textbf{k'}-\frac{1}{2}\textbf{q}_{2}}
     +   \nonumber \\
 &+& T_{\textbf{k}+\frac{1}{2}\textbf{q}_{1},\textbf{k'}+\frac{1}{2}\textbf{q}_{2}}
G^{*-1} _{\textbf{k'}-\frac{1}{2}\textbf{q}_{2}}
\delta_{\textbf{k}-\frac{1}{2}\textbf{q}_{1},
\textbf{k'}-\frac{1}{2}\textbf{q}_{2}}   \nonumber \\
&+&T^{*}_{\textbf{k}-\frac{1}{2}\textbf{q}_{1},\textbf{k'}-\frac{1}{2}\textbf{q}_{2}}
G^{-1} _{\textbf{k'}+\frac{1}{2}\textbf{q}_{2}}
\delta_{\textbf{k}+\frac{1}{2}\textbf{q}_{1},
\textbf{k'}+\frac{1}{2}\textbf{q}_{2}}\, , \nonumber
\end{eqnarray}
where we abbreviated $G(\omega,\textbf{k})=G_{\textbf{k}}$.
 $\rho KS$ is  negligible for a distant source
 and $\rho K\rho S$ just
 slightly modifies the diffuse background $\rho$. The
terms $ \rho S + K\rho S$ dominate if the object is less than a
mean free path separated from the receiver, and give us,
\begin{eqnarray}\label{ob}
   \hspace{-2cm}&& \left< \Psi_{\textbf{k}+\frac{1}{2}\textbf{q}_{1}} \Psi^*_{\textbf{k}-\frac{1}{2}\textbf{q}_{1}}\right>
    =G_{\textbf{k}+\frac{1}{2}\textbf{q}_{1}} G^{*}_{\textbf{k}-\frac{1}{2}\textbf{q}_{1}}
      \nonumber \\
     &&\sum_{\textbf{k'}}  \left[ \delta_{\textbf{kk'}}+ K_{\textbf{kk'}}(\textbf{q}_{1},
    \textbf{q}_{2})\right]
    \frac{-1}{\omega} \textrm{Im}\, G_{\textbf{k'}}
    \rho(\textbf{q}_{2})S(\omega)\, .
\end{eqnarray}
The Ward identity \cite{mahan},
\begin{eqnarray}\label{wi}
&& T_{\textbf{p}+\frac{1}{2}\textbf{q},
\textbf{p'}+\frac{1}{2}\textbf{q}}
-T^{*}_{\textbf{p}-\frac{1}{2}\textbf{q},
\textbf{p'}-\frac{1}{2}\textbf{q}} = \nonumber \\
&& \  \ \sum_{\textbf{k}} T_{\textbf{p}+\frac{1}{2}\textbf{q},
\textbf{k}+\frac{1}{2}\textbf{q}}\left[G_{\textbf{k}+\frac{1}{2}\textbf{q}}-
G^{*}_{\textbf{k}-\frac{1}{2}\textbf{q}}\right]
T^{*}_{\textbf{p'}-\frac{1}{2}\textbf{q},
\textbf{k}-\frac{1}{2}\textbf{q}}\, . \nonumber
\end{eqnarray}
expressing flux conservation (assumption 2) simplifies this
expression. The singular diffusion pole (assumption 1) imposes
$q_{2}\approx 0$ in all T-matrices and Green functions. The
assertion~(\ref{object}) follows after Fourier transformations
with respect to $\omega$,$\textbf{k}$, $\textbf{q}_{1}$ and
$\textbf{q}_{2}$.

We will finally make the link of Eq.~(\ref{corr2}) with
time-reversal (TR) experiments \cite{finkall} and coherent
backscattering (CBS). Consider a source $s(t)$ at point
$\textbf{r}_{S}$ in an infinite random medium. At point
$\textbf{r}_{T}$  a TR process is carried out between times
$T_{1}$ and $T_{2}$. The TR signal $\Psi(\textbf{r},\tau )$
collected at $\textbf{r}_{D}$ was given by De Rosny \emph{etal}.
\cite{julienthese} and Papanicolaou \emph{etal}. \cite{george2};
\begin{equation}\label{tr}
\Psi(\textbf{r}_{D},\tau + 2T_{2})= \int_{T_{1}}^{T_{2}}
\textrm{d}t\, G(\textbf{r}_{S}, \textbf{r}_{T},\tau+
t)G(\textbf{r}_{T}, \textbf{r}_{D},t ) \otimes s(t,\textbf{r}_{S})
\end{equation}
If we agree to time-reverse the whole signal ($T_{1}$  smaller
than the first arrival time and $T_{2}$ deep inside the coda), a
Fourier transformation gives us,
\begin{eqnarray}\label{interTR}
\Psi(\textbf{r}_{D},\tau +  2T_{2} ) =  \int_{-\infty}^{\infty}
\frac{\textrm{d} \omega}{2\pi}\, \textrm{e}^{-i\omega\tau}\,
G(\textbf{r}_{S}, \textbf{r}_{T},\omega)G^*(\textbf{r}_{T},
\textbf{r}_{D}, \omega )\nonumber \\
 \times  s(\omega) \, .
\end{eqnarray}
The similarity to $\Phi_{\textbf{r}}(\textbf{x},\tau)$ (with
$\textbf{r}=\textbf{r}_{S}-\textbf{r}_{T}$ and
$\textbf{x}=\textbf{r}_{S}-\textbf{r}_{D}$) given by
Eq.~(\ref{inter1}) expressing diffuse correlations is striking.
Thus Eq.~(\ref{nodisp}) explains the auto-focalization in
space-time, whereas the TR signal $\Psi(\textbf{r},\tau )$ is very
stable against statistical fluctuations provided it has a
bandwidth much larger than the Thouless frequency. In
Ref.~\cite{derode95} the ratio equals $B/\Omega_{\textrm{th}}
\approx 600$.

The link between TR and  CBS can be established by applying
formula~(\ref{interTR}) to a disordered half space with a TR
machine hidden inside at a depth $z_{T}$ and a source at a large
distance $a$ outside (Figure 1). We assume $z_{T}\ll a$.
Equation~(\ref{interTR}) predicts that the TR signal at a distance
$x$ from the source is,
\begin{eqnarray}\label{C}
\Psi(\textbf{x},\tau+2T_{2}) &=&  s(\tau)\,
 \int
\textrm{d}^{2}\textbf{r}_{\parallel} \, e^{ik\textbf{x}\cdot
\textbf{r}_{\parallel}/a}\rho (\ell,z_{T}, \textbf{r}_{\parallel}
) \,
 \nonumber \\
&\sim & s(\tau)\, \frac{\ell^*}{z_{T}}
\exp\left(-\frac{z_{T}k|x|}{a} \right)
\end{eqnarray}
where we have inserted the diffuse energy propagator
$\rho(z,z',\textbf{r}_{\parallel}-\textbf{r}'_{\parallel})$ for
the 3D half space \cite{akker}. This makes the line shape of the
spatial auto-focalization in TR \emph{equal} to the one of CBS
from a half space, exhibiting the well-known angular triangular
cusp caused by very long time-reversed wave paths \cite{akker}.
The angular resolution $\Delta x/a \approx \lambda/z_{T} $ nicely
illustrates that the diffusion process creates an effective
aperture of size $z_{T}$ that increases the quality of spatial
focalization, \emph{independent} of $\ell^*$ \cite{derode95}. The
peak signal is \emph{stable} against statistical fluctuations if
$B\gg 2D/z_{T}^{2}$.

In conclusion, we have shown that field-correlations of diffuse
waves can be used to retrieve  ballistic waves between two points
in space-time. This method is stable against the unknown
statistical fluctuations  that cause the field to be diffuse,
provided that the operating bandwidth is much larger than the
Thouless frequency. We have discussed the possible temporal
asymmetry of the field correlations,  the role of the power
spectrum of the source, as well as the fundamental relation to
time-reversal and coherent backscattering. The method facilitates
a passive way of imaging that might find an application in
seismology, where active sources are expensive or unpredictable.

This work has been accomplished thanks to pertinent discussions
 with
Michel Campillo, Mathias Fink, Ludovic Margerin, Roger Maynard,
Julien de Rosny, John Scales, Sergey Skipetrov, Arnaud Tourin, and
Richard Weaver. We acknowledge support by the French Ministry of
Research ACI 2066, the GDR 2253 IMCODE of the CNRS, and NSF/CNRS
contract 14872.

\begin{figure}
\includegraphics[height=0.4\hsize,angle=0]{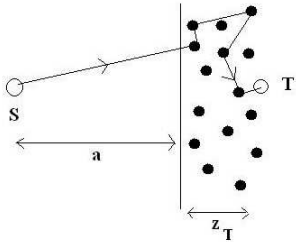}
\caption{\label{fig:one} A set-up that establishes the link
between coherent backscattering and time-reversal. A source
\textbf{S} emits a signal that is received by a time-reversal
machine \textbf{T} inside a disordered half-space. The signal that
is sent back auto-focalizes in space and time near \textbf{S} with
the spatial line-shape of coherent backscattering. }
\end{figure}

\end{document}